\documentclass[aps,prl,twocolumn,superscriptaddress,showpacs,preprintnumbers]{revtex4}
\usepackage[pdftex]{graphicx,color}
\usepackage{hyperref}
\input{epsf}
\usepackage{amsmath,amssymb}
\usepackage[dvips,letterpaper,text={6.5in,9in}]{geometry}
\usepackage{fancyhdr}
\usepackage{verbatim}

\newcommand\ltap{\
  \raise.3ex\hbox{$<$\kern-.75em\lower1ex\hbox{$\sim$}}\ }
\newcommand\gtap{\
  \raise.3ex\hbox{$>$\kern-.75em\lower1ex\hbox{$\sim$}}\ }

\newcommand\simge{\mathrel{%
   \rlap{\raise 0.511ex \hbox{$>$}}{\lower 0.511ex \hbox{$\sim$}}}}
\newcommand\simle{\mathrel{
   \rlap{\raise 0.511ex \hbox{$<$}}{\lower 0.511ex \hbox{$\sim$}}}}

\newcommand{\slashchar}[1]%
        {\kern .25em\raise.18ex\hbox{$/$}\kern-.75em #1}
\def\lsim{\mathrel{\raise.3ex\hbox{$<$\kern-.75em\lower1ex\hbox{$\sim$}}}}
\def\gsim{\mathrel{\raise.3ex\hbox{$>$\kern-.75em\lower1ex\hbox{$\sim$}}}}
\newcommand{\bs}{\boldsymbol}

\newcommand\CL{{\cal L}}

\newcommand\CO{{\cal O}}

\newcommand\be{\begin{equation}}
\newcommand\ee{\end{equation}}
\newcommand\bea{\begin{eqnarray}}
\newcommand\eea{\end{eqnarray}}
\newcommand\ba{\begin{array}}
\newcommand\ea{\end{array}}

\newcommand{\thw}{\ensuremath{\theta_W}}

\newcommand\ra{\rightarrow}

\newcommand\gev{{\rm GeV}}
\newcommand\tev{{\rm TeV}}

\newcommand\pb{{\rm pb}}

\newcommand\fb{{\rm fb}}
\newcommand\ifb{{\rm fb}^{-1}}

\newcommand\etmiss{\slashchar{E}_T}

\newcommand\ellm{\ell^-}
\newcommand\ellpm{\ell^\pm}
\newcommand\ellp{\ell^+}

\newcommand\Ntc{N_{TC}}

\newcommand\Ltc{\Lambda_{TC}}

\newcommand\tom{\omega_{T}}
\newcommand\tro{\rho_{T}}

\newcommand\ta{a_T}

\newcommand\tapm{a_T^\pm}

\newcommand\taz{a_T^0}

\newcommand\tropm{\rho_{T}^\pm}

\newcommand\troz{\rho_{T}^0}

\newcommand\tpi{\pi_T}
\newcommand\tpipm{\pi_T^\pm}
\newcommand\tpimp{\pi_T^\mp}
\newcommand\tpip{\pi_T^+}

\newcommand\tpiz{\pi_T^0}

\newcommand\jets{{\rm jets}}

\newcommand\Mjj{M_{jj}}

\newcommand\MWjj{M_{Wjj}}

\begin{document}
 
 \preprint{FERMILAB-Pub-11-165-T}

\title{Technicolor Explanation for the CDF $Wjj$ Excess}

\author{Estia J.~Eichten}
 \email[E-mail:]{eichten@fnal.gov}
 \affiliation{Theoretical Physics Department, Fermi National Accelerator
   Laboratory\\ P.O.\ Box 500, Batavia, Illinois 60510}
\author{Kenneth Lane}
 \email[E-mail:]{lane@physics.bu.edu}
 \affiliation{Department of Physics, Boston University\\ 590 Commonwealth
   Avenue, Boston, Massachusetts 02215} 
\author{Adam Martin}
 \email[E-mail:]{aomartin@fnal.gov}
 \affiliation{Theoretical Physics Department, Fermi National Accelerator
   Laboratory\\ P.O.\ Box 500, Batavia, Illinois 60510}

\begin{abstract}

  We propose that the $3.2\,\sigma$ excess at $\sim 150\,\gev$ in the dijet
  mass spectrum of $W+\jets$ reported by CDF is the technipion $\tpi$ of
  low-scale technicolor. Its relatively large cross section is due to
  production of a narrow $Wjj$ resonance, the technirho, which decays to
  $W\tpi$. We discuss ways to enhance and strengthen the technicolor
  hypothesis and suggest companion searches at the Tevatron and LHC.

\end{abstract}
\maketitle


\newpage

{\em {1. \underbar {Introduction}}} \,\, The CDF Collaboration has reported a
surprising excess at $\Mjj \simeq 150\,\gev$ in the dijet mass distribution
of $W + \jets$ events. Fitting the excess to a Gaussian, CDF estimated its
production rate to be $\sim 4\,\pb$. This is 300~times the standard model
Higgs rate $\sigma(\bar pp \ra WH)B(H\ra \bar bb)$. The Gaussian fit is
consistent with a zero-width resonance. Its significance, for a search window
of 120--$200\,\gev$ and including systematic uncertainties, is
$3.2\,\sigma$~\cite{Aaltonen:2011mk}.

In our view the most plausible new-physics explanation of this excess is
resonant production and decay of bound states of technicolor (TC), a new
strong interaction at $\Ltc \sim$ several $100\,\gev$ of massless
technifermions~\cite{Weinberg:1979bn,Susskind:1978ms,Hill:2002ap,
  Lane:2002wv}. These technifermions are assumed to belong to complex
representations of the TC gauge group and transform as quarks and leptons do
under electroweak (EW) $SU(2)\otimes U(1)$. Then, the spontaneous breaking of
their chiral symmetry breaks EW symmetry down to electromagnetic $U(1)$ with
a massless photon and $M_W/M_Z\cos\thw = 1 + \CO(\alpha)$. We propose that
the dijet resonance is the lightest pseudo-Goldstone isovector technipion
($\tpi$) of the low-scale technicolor scenario. The immediate consequence of
this hypothesis is a narrow $I=1$ technirho ($\tro$) resonance in the $Wjj$
channel. This accounts for the large $W\tpi$ production rate.

In this Letter we show that a $\tro$ of mass $\simeq 290\,\gev$ decaying into
$W$ plus $\tpi$ of $160\,\gev$ accounts for the CDF dijet excess. The $\tro$
signal sits near the peak of the $\MWjj$ distribution and will be less
obvious than $\tpi \ra jj$. We suggest ways to enhance this signal and tests
of the $\tro$'s presence: (1) The $\tro$'s narrowness will be reflected in
$Q = \MWjj - \Mjj - M_W$~\cite{Eichten:1997yq,Aaltonen:2009jb}.
The $\Mjj$ bins near $M_{\tpi}$ will exhibit a sharp increase over background
for $ Q \simeq Q^* = M_{\tro} - M_{\tpi} - M_W$. (2) The $\tro \ra W\tpi$
angular distribution in the $\tro$ frame will be approximately
$\sin^2\theta$, indicative of the signal's technicolor origin.  We propose
further tests of the technicolor hypothesis, including other resonantly
produced states which can be discovered at the Tevatron and LHC.

Low-scale technicolor (LSTC) is a phenomenology based on walking
technicolor~\cite{Holdom:1981rm, Appelquist:1986an,Yamawaki:1986zg,
  Akiba:1986rr}. The TC gauge coupling must run very slowly for 100s of TeV
above $\Ltc$ so that extended technicolor (ETC) can generate sizable quark
and lepton masses~\footnote{Except for the top quark mass, which requires additional dynamics such as topcolor~\cite{Hill:1994hp}.} while suppressing flavor-changing neutral current
interactions~\cite{Eichten:1979ah}. This may be achieved if technifermions
belong to higher-dimensional representations of the TC gauge group.  Then,
the constraints of Ref.~\cite{Eichten:1979ah} on the number of ETC-fermion
representations imply technifermions in the fundamental representation as
well. Thus, there are technifermions whose technipions' decay constant $F_1^2
\ll F_\pi^2 = (246\,\gev)^2$~\cite{Lane:1989ej}. Bound states of these
technifermions will have masses well below a TeV --- greater than the limit
$M_{\tro} \simge 250\,\gev$~\cite{Abazov:2006iq, Aaltonen:2009jb} and
probably less than the 600--700~GeV at which ``low-scale'' TC ceases to make
sense. Technifermions in complex TC representations imply a quarkonium-like
spectrum of mesons. The most accessible are the lightest technivectors, $V_T
= \tro(I^G J^{PC} = 1^+1^{--})$, $\tom(0^-1^{--})$ and $\ta(1^-1^{++})$;
these are produced as $s$-channel resonances in the Drell-Yan process in
hadron colliders.  Technipions $\tpi(1^-0^{-+})$ are accessed in $V_T$
decays. A central assumption of LSTC is that these technihadrons may be
treated in isolation, without significant mixing or other interference from
higher-mass states.  Also, we expect that (1) the lightest technifermions are
$SU(3)$-color singlets, (2) isospin violation is small for $V_T$ and $\tpi$,
(3) $M_{\tom} \cong M_{\tro}$, and (4) $M_{\ta}$ is not far above $M_{\tro}$.
An extensive discussion of LSTC, including these points and precision
electroweak constraints, is given in Ref.~\cite{Lane:2009ct}.

Walking technicolor has another important consequence: it enhances $M_{\tpi}$
relative to $M_{\tro}$ so that the all-$\tpi$ decay channels of the $V_T$
likely are closed~\cite{Lane:1989ej}. Principal $V_T$-decay modes are
$W\tpi$, $Z\tpi$, $\gamma \tpi$, a pair of EW bosons (including one photon),
and fermion-antifermion pairs~\cite{Lane:2002sm, Eichten:2007sx,Lane:2009ct}.
If allowed by isospin, parity and angular momentum, $V_T$ decays to one or
more weak bosons involve longitudinally-polarized $W_L/Z_L$, the technipions
absorbed via the Higgs mechanism. These nominally strong decays are
suppressed by powers of $\sin\chi = F_1/F_\pi \ll 1$. Decays to
transversely-polarized $\gamma,W_\perp,Z_\perp$ are suppressed by
$g,g'$. Thus, the $V_T$ are {\em very} narrow, $\Gamma(V_T) \simle 1\,\gev$.
These decays provide striking signatures, visible above backgrounds within a
limited mass range at the Tevatron and probably up to 600--700~GeV at the
LHC~\cite{Brooijmans:2008se, Brooijmans:2010tn}.

{\em {2. \underbar {The new dijet resonance at the Tevatron}}} \,\, Previous
$\tro \ra W\tpi$ searches at the Tevatron focused on final states with $W \ra
\ell\nu_\ell$ and $\tpi \ra \bar q q$ where one or both quarks was a
tagged~$b$. This was advocated in Ref.~\cite{Eichten:1997yq} because $\tpi$
couplings to standard-model fermions are induced by ETC interactions and are,
naively, expected to be largest for the heaviest fermions. Thus, $\tpip \ra
\bar b c$, $\bar b u$ and $\tpiz \ra \bar b b$ has been assumed, at least for
$M_{\tpi} \simle m_t$. While reasonable for $\tpiz$ decays, it is
questionable for $\tpipm$ because CKM-like angles may suppress $\bar b q$.
This is important because the {\em inclusive} $\sigma(\bar u u,\bar d d \ra
\troz) \simeq 1.6\times\, \sigma(\bar d u,\bar u d \ra \tropm)$ at the Tevatron. If
$\tpip \ra \bar b q$ is turned off in the default model of $\tpi$ decays used
here~\cite{Lane:2002sm}, up to 40\% of the $\tro \ra W\tpi \ra Wjj$ signal is
vetoed by a $b$-tag . It is notable, therefore, that the CDF observation did
not require $b$-tagged jets~\cite{Aaltonen:2011mk}.

At first, it seems unlikely that $\tro \ra W\tpi$ could be found in untagged
dijets because of the large $W+\jets$ background. However,
Ref.~\cite{Mrenna:1999xj} studied $\tro\ra W \tpiz$ without flavor-tagging
and showed that a $\tpi \ra jj$ signal could be extracted. Recently, strong
$W/Z \ra jj$ signals have been observed in $WW/WZ$ production at the
Tevatron~\cite{Aaltonen:2009vh, Aaltonen:2010rq}. So, heavier dijet states
resonantly produced with $W/Z/\gamma$ may indeed be discoverable at the
Tevatron.

The CDF dijet excess was enhanced by requiring $p_T(jj) >40\,\gev$~\cite{Aaltonen:2011mk}.
Such a cut was proposed in Ref.~\cite{Eichten:1997yq}. There it was
emphasized that the small $Q$-value in $\tro \ra W\tpi$ and the fact that
the $\tro$ is approximately at rest in the Tevatron lab frame cause the
$\tpi$ to be emitted with limited $p_T$ and its decay jets to be roughly
back-to-back in~$\phi$.

{\em {3. \underbar{Simulating $\rho_T \ra W\tpi$}}}\,\, {\sc Pythia}~6.4 is
used throughout to generate the $\rho_T \ra W\tpi$
signal~\cite{Sjostrand:2006za}. It employs the default $\tpi$-decay model of
Ref.~\cite{Lane:2002sm} in which $\tpip \ra \bar b q$ is unhindered. The
input masses are $(M_{\tro},\,M_{\tpi}) = (290,\,160)\,\gev$. This $M_{\tpi}$
gives a peak in the simulated $\Mjj$ distribution near
$150\,\gev$~\footnote{Other relevant LSTC masses are $M_{\tom} = M_{\tro}$;
  $M_{\ta} = 1.1 M_{\tro} = 320\,\gev$; and $M_{V_i,A_i}$ which appear in
  dimension-five operators for $V_T$ decays to transverse EW
  boson~\cite{Lane:2002sm,Eichten:2007sx}; we take them equal to $M_{\tro}$.
  Other LSTC parameters are $\sin\chi = 1/3$, $Q_U = Q_D + 1 = 1$, and $\Ntc
  = 4$.}. This parameter choice is close to Case~2b of Contribution~8 in
Ref.~\cite{Brooijmans:2010tn}.

The signal cross sections ({\em including} $B(\tpiz \ra \bar q q) \simeq
0.90$, $B(\tpipm \ra \bar q q') \simeq 0.95$, and $B(W \ra \ellpm\nu_\ell) =
0.21$) are $\sigma(W^\pm\tpimp) = 310\,\fb$ and $\sigma(W^\pm\tpiz) =
175\,\fb$~\footnote{No $K$-factor has been used in any of our signal and
  background calculations.}. Only 20-30\% of these cross sections come from
the $320\,\gev$ $a_T \ra W_\perp \tpi$. If $M_{\ta} = 293\,\gev$, they
increase slightly to $335\,\fb$ and $205\,\fb$. If $\tpip \ra \bar b q$ is
suppressed, then $\sigma(W^\pm\tpimp) = 110\,\fb$, a decrease of 2/3, for a
total $Wjj$ signal of $285\,\fb$.

Backgrounds come from standard model $W/Z + \jets$, including $b,\,c$-jets,
$WW/WZ$, $t\bar t$, and multijet QCD. The last two amount to $\sim 10\%$ at
the Tevatron and we neglect them. The others are generated at parton
level with ALPGENv13~\cite{Mangano:2002ea} and fed into {\sc Pythia} for
showering and hadronization. The {\sc Pythia} particle-level output is
distributed into calorimeter cells of size $\Delta \eta \times \Delta \phi =
0.1\times 0.1$.
After isolated leptons (and photons) are removed, all remaining cells with
$E_T> 1\,\gev$ are used for jet-finding. Jets are defined using a midpoint
cone algorithm with $R = 0.4$. For simplicity, we did not smear calorimeter
energies; this does not significantly broaden our $\Mjj$ resolution near
$M_{\tpi}$.

%
\begin{figure}[!t]
 \begin{center}
   \includegraphics[width=2.60in,
   height=2.00in]{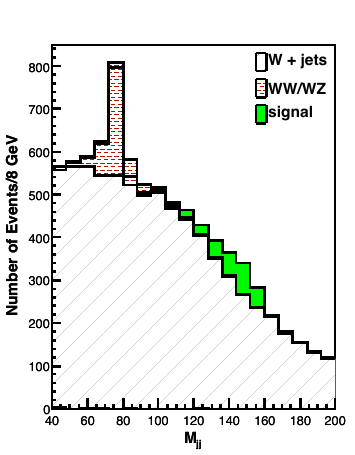}
   \includegraphics[width=2.60in,
   height=2.00in]{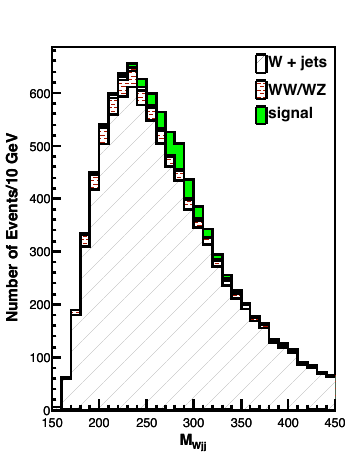}
   \caption{The $\Mjj$ and $\MWjj$ distributions in $\bar pp$ collisions at
     $1.96\,\tev$ for LSTC with $M_{\tro} = 290\,\gev$, $M_{\tpi} =
     160\,\gev$ and $\int \CL dt = 4.3\,\ifb$. Only the CDF cuts described in
     the text are used.\label{fig:CDF_mass_dists}}
 \end{center}
 \end{figure}

 In extracting the $\tpi$ and $\tro$ signals, we first adopted the cuts used
 by CDF~\cite{Aaltonen:2011mk},\footnote{The CDF cuts are: exactly one
   lepton, $\ell = e,\mu$, with $p_T > 20\,\gev$ and $|\eta| < 1.0$; exactly
   two jets with $p_T > 30\,\gev$ and $|\eta| < 2.4$; $\Delta R(\ell, j) >
   0.52$; $p_T(j_1 j_2) > 40\,\gev$; $\etmiss > 25\,\gev$; $M_T(W) >
   30\,\gev$; $|\Delta\eta(j_1j_2)| < 2.5$; $|\Delta\phi(\bs{\etmiss},j_1)| >
   0.4$.}. Our results are in Fig.~\ref{fig:CDF_mass_dists}. The data
 correspond to $\int\CL dt = 4.3\,\ifb$. They reproduce the shape and
 normalization of CDF's $\Mjj$~\cite{Aaltonen:2011mk} and
 $\MWjj$~\cite{Cavaliere:2010zz} distributions (except that not smearing
 calorimeter energies does make our $W\ra jj$ signal a narrow spike). We
 obtain $S/B = 250/1595$ for the dijet signal in the five bins in
 120--160~GeV. We find this agreement with CDF's measurement remarkable. Our
 model inputs are standard defaults, chosen only to match the dijet resonance
 position and the small $Q$-value of $\tro \ra W\tpi$. The $\tro$ resonance
 is near the peak of the $\MWjj$ distribution~\footnote{The quadratic
   ambiguity in the $W$ reconstruction was resolved by choosing the solution
   with the smaller $p_z(\nu)$.}. For the six bins in 240--300~GeV, we obtain
 $S/B = 235/3390$.

 We then augmented the CDF cuts to enhance the signals. CDF required exactly
 two jets. We achieved greater acceptance and a modest sharpening of the
 dijet peak by combining a third jet with one of the two leading jets if it
 was within $\Delta R = 1.5$ of either of them. We enhanced the $\tpi$ and,
 especially, the $\tro$ signals by imposing topological cuts taking
 advantage of the $\tro \ra W\tpi$ kinematics~\cite{Eichten:1997yq}:
 (1) $\Delta \phi(j_1j_2) > 1.75$ and (2) $p_T(W) =|{\bs p}_T(\ell) + {\bs
   p}_T(\nu_\ell)| > 60\,\gev$. The improvements seen in
 Fig.~\ref{fig:ELM_CDF_mass_dists} are significant. We obtain $S/B = 200/800$
 for $\tpi \ra jj$ and $S/B = 215/1215$ for $\tro \ra Wjj$. Extracting the
 $\tro$ signal will require confidence in the background shape.

\begin{figure}[!t]
 \begin{center}
   \includegraphics[width=2.60in,
   height=2.00in]{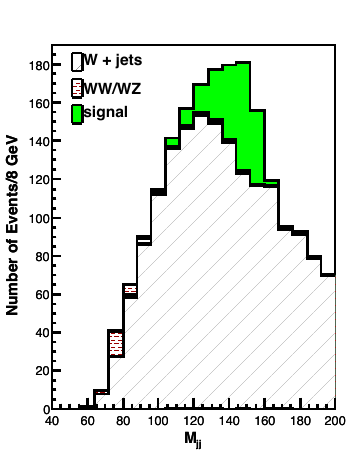}
   \includegraphics[width=2.60in,
   height=2.00in]{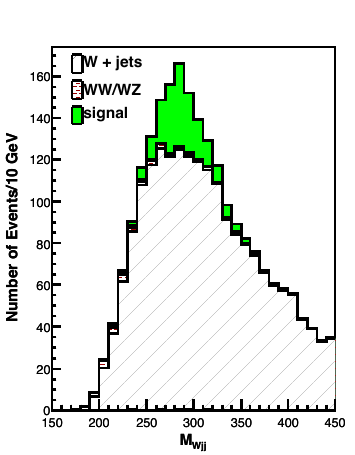}
   \caption{The $\Mjj$ and $\MWjj$ distributions in $\bar pp$ collisions at
     $1.96\,\tev$ for LSTC with $M_{\tro} = 290\,\gev$, $M_{\tpi} =
     160\,\gev$ and $\int \CL dt = 4.3\,\ifb$. CDF cuts augmented with ours
     described in the text are used.\label{fig:ELM_CDF_mass_dists}}
 \end{center}
 \end{figure}
\begin{figure}[!t]
 \begin{center}
\includegraphics[width=2.60in, height=2.00in]{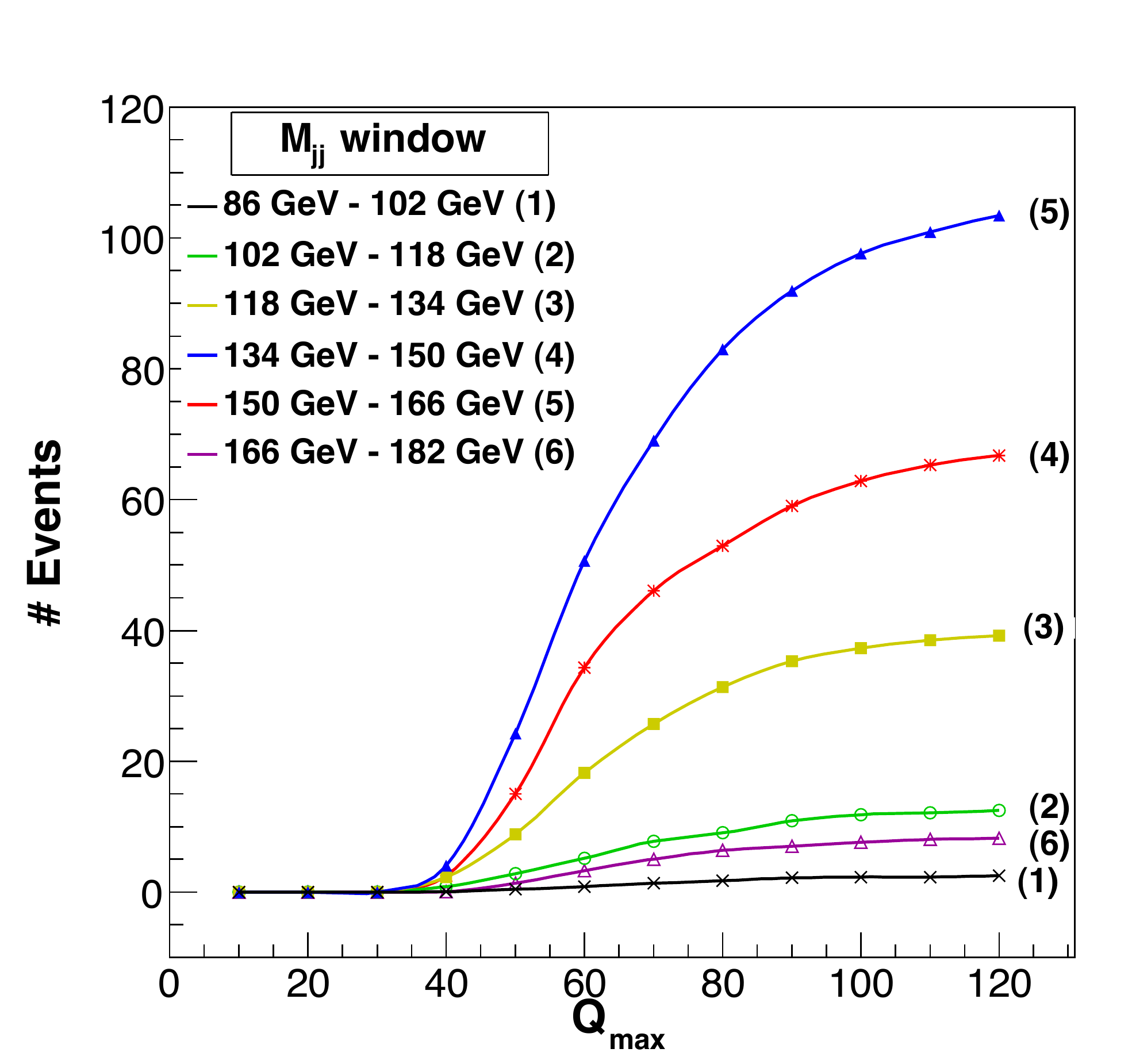}
\includegraphics[width=2.60in, height=2.00in]{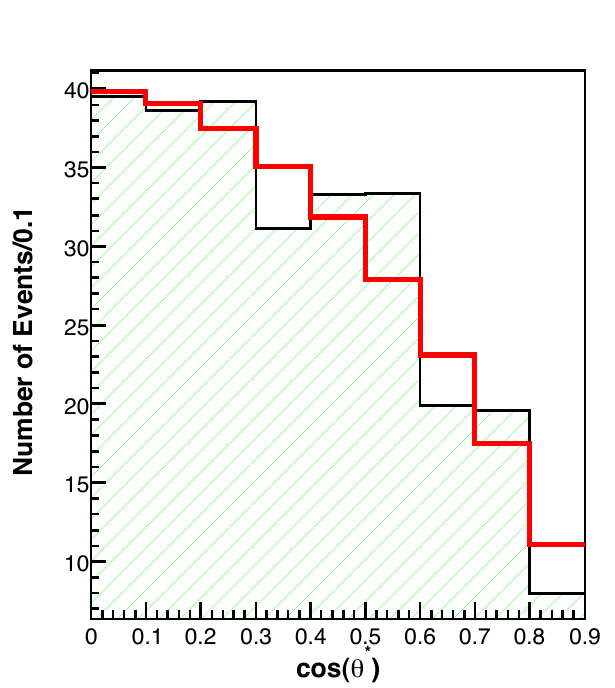}
\caption{Top: $\Delta N(\Mjj)$ vs.~$Q_{\rm max}$ as described in the text
  for the indicated $\Mjj$ bins. Bottom: The background-subtracted $W$-dijet
  angular distribution, compared to $\sin^2\theta$ (red).
  \label{fig:Q_angular}}
 \end{center}
 \end{figure}
%



 In addition to the $jj$ and $Wjj$ resonances, the $Q$-value and the
 $\tro$-decay angular distribution are indicative of resonant production of
 $W\tpi$. The resolution in $Q = M_{Wjj} - M_{jj} - M_W$ is better than in
 $M_{jj}$ and $M_{Wjj}$ alone because jet measurement errors partially
 cancel. This is seen in Fig.~\ref{fig:Q_angular} where we plot $\Delta
 N(M_{jj}) = N_{\rm observed}(M_{jj})-N_{\rm expected}(M_{jj}) =
 N_{S+B}(M_{jj}) - N_B(M_{jj})$ for $Q \le Q_{\rm max}$ vs.~$Q_{\rm max}$ for
 six 16-GeV $\Mjj$ bins between 86~and $182\,\gev$. The sudden increase at
 $Q_{\rm max} \simeq 50\,\gev$ in the three signal bins is clear.
 
 The decay $\tro \ra W\tpi$ is dominated by $ W_L\tpi$. Therefore, the
 angular distribution of $q \bar q \ra \tro \ra W\tpi$ is approximately
 $\sin^2\theta$, where $\theta$ is the angle between the incoming quark and
 the outgoing $W$ in the $\tro$ frame~\cite{Eichten:2007sx}. The backgrounds
 are forward-backward peaked. We required $p_T(W) > 40\,\gev$, fit the
 background in $250 < M_{Wjj} < 300\,\gev$ with a quartic in $\cos\theta$,
 and subtracted it from the total. (In reality, of course, one would use
 sidebands.) The prediction in Fig.~\ref{fig:Q_angular} matches the
 normalized $\sin^2\theta$ well. Verification of this would strongly support
 the TC origin of the signal.

{\em {4. \underbar{Other LSTC tests at the Tevatron and LHC}}}\,\,

\noindent 1) It is important to find the $\tom$ and $\ta$ states, expected to
be close to $\tro$, near $300\,\gev$. At the Tevatron, the largest production
rates involve $\tom \ra \gamma\tpiz$ and $\tapm \ra \gamma \tpipm$. For our
input parameters, these are $80\,\fb$ and $185\,\fb$, respectively. Their
existence, masses and production rates critically test the technifermions' TC
representation structure and the strength of the dimension-five operators
inducing these decays. In addition, recent papers from
D\O~\cite{Abazov:2010ti} and CDF~\cite{Aaltonen:2011xp} suggest that the
$e^+e^-$ channel is promising. The excess (signal) cross sections for our
parameters are $\sigma(\tom,\troz \ra e^+e^-) = 12\,\fb$ and $\sigma(\taz \ra
e^+e^-) = 7\,\fb$.

\noindent 2) Finding these LSTC signatures at the LHC is complicated by $\bar
tt$ and other multijet backgrounds. The likely discovery and study channels
at the LHC are the nonhadronic final states of $\tropm \ra W^\pm Z^0$;
$\tropm, \tapm \ra \gamma W^\pm$, and $\troz,\tom,\taz \ra
\ellp\ellm$~\cite{Brooijmans:2008se, Brooijmans:2010tn}. The dilepton
channel may well be the earliest target of opportunity.

\noindent 3) The $b$ and $\tau$-fractions of $\tpi$ decays should be
determined as well as possible. They probe the ETC couplings of quarks and
leptons to technifermions, a key part of the flavor physics of dynamical
electroweak symmetry breaking~\cite{Eichten:1979ah}.

If experiments at the Tevatron and LHC reveal a spectrum resembling these
predictions, it could well be that low-scale technicolor is the ``Rosetta
Stone'' of electroweak symmetry breaking. For it will then be possible to
know its dynamical origin and discern the character of its basic
constituents, the technifermions. The masses and quantum numbers of their
bound states will provide stringent experimental benchmarks for the
theoretical studies of the strong dynamics of walking technicolor just now
getting started, see e.g.~\cite{ Appelquist:2010xv}.

{\em \underbar{Acknowledgments}} We are grateful to K.~Black, T.~Bose,
J.~Butler, J.~Campbell, K.~Ellis, W.~Giele, C.~T.~Hill, E.~Pilon and
J.~Womersley for valuable conversations and advice. This work was supported
by Fermilab operated by Fermi Research Alliance, LLC, U.S.~Department of
Energy Contract~DE-AC02-07CH11359 (EE and AM) and in part by the
U.S.~Department of Energy under Grant~DE-FG02-91ER40676~(KL). KL's research
was also supported in part by Laboratoire d'Annecy-le-Vieux de Physique
Theorique (LAPTH) and he thanks LAPTH for its hospitality.

{\em {\underbar{Note added in proof}}} -- An important corroboration of the $\rho_T \rightarrow W^{\pm}\,\pi_T \rightarrow \ell^{\pm}\nu_{\ell}\,jj$ signal
is its isospin partner (suppressed by phase space and branching ratios) $\rho_T \rightarrow Z^{0}\,\pi^{\pm}_T \rightarrow \ell^{+}\,\ell^-\,jj$.  We predict cross sections of $38\, \text{fb}$ at the Tevatron and $155\, \text{fb}$ at the 7-TeV LHC for $\ell = e$ and $\mu$.


\bibliography{TC_at_TeV_S}
\bibliographystyle{utcaps}
\end{document}